\documentclass[10pt]{bmc_article}

\usepackage{cite} 
\usepackage{url}  
\usepackage{ifthen}  
\usepackage{multicol}   
\usepackage[utf8]{inputenc} 
\def\includegraphics{}

\newboolean{publ}

\newcommand{\Starr}{{\em Starr} }
\newcommand{\Ringo}{{\em Ringo} }

\newenvironment{bmcformat}{\begin{raggedright}\baselineskip20pt\sloppy\setboolean{publ}{false}}{\end{raggedright}\baselineskip20pt\sloppy}

\begin{document}
\begin{bmcformat}


\title{Starr: Simple Tiling ARRay analysis of Affymetrix ChIP-chip data}
 

\author{
        Benedikt Zacher\correspondingauthor$^1$ 
         \email{Benedikt Zacher - zacher@lmb.uni-muenchen.de}%
      \and
         Achim Tresch$^1$
       \email{Achim Tresch\correspondingauthor - tresch@lmb.uni-muenchen.de}%
      }


\address{%
    \iid(1) Gene Center, Ludwig-Maximilians-University of Munich, Feodor-Lynen-Str. 25, D-81377 Munich, Germany
}%

\maketitle


\begin{abstract}
        \paragraph*{Background:} Chromatin immunoprecipitation combined with DNA microarrays (ChIP-chip) is an assay for DNA-protein-binding or post-translational chromatin/histone modifications. As with all high-throughput technologies, it requires a thorough bioinformatic processing of the data for which there is no standard yet. The primary goal is the reliable identification and localization of genomic regions that bind a specific protein. The second step comprises comparison of binding profiles of functionally related proteins, or of binding profiles of the same protein in different genetic backgrounds or environmental conditions. Ultimately, one would like to gain a mechanistic understanding of the effects of DNA binding events on gene expression.
           
        \paragraph*{Results:} We present a free, open-source {\bf R} package \Starr that, in combination with the package \Ringo, facilitates the comparative analysis of ChIP-chip data across experiments and across different microarray platforms. Core features are data import, quality assessment, normalization and visualization of the data, and the detection of ChIP-enriched genomic regions. The use of common Bioconductor classes ensures the compatibility with other {\bf R} packages.
        
        \paragraph*{Conclusion:} \Starr is an {\bf R} package that enables flexible analysis of a wide range of ChIP-chip experiments, in particular for Affymetrix data. Most importantly, \Starr provides methods for integration of complementary genomics data, e.g., it enables systematic investigation of the relation between gene expression and DNA binding.

\end{abstract}

\ifthenelse{\boolean{publ}}{\begin{multicols}{2}}{}


\section*{Background}
ChIP-chip is a technique for identifying Protein-DNA interactions. For this purpose, the chromatin is immunoprecipitated with an antibody to the protein of interest and the fragmented, protein-bound DNA is analyzed with tiling arrays \cite{chipchip}. Before the results can be analyzed, some bioinformatics methods must be applied to ensure the quality of the experiments and preprocess the data. \\

Here we present the open-source software package \Starr, which is available as part of the opensource bioconductor project \cite{bioconductor}. It is an extension package for the programming language
and statistical environment R \cite{R}. \Starr facilitates the analysis of ChIP-chip data, in particular it supports experiments that have been performed on the Affymetrix\textsuperscript{\texttrademark}
 platform. Its functionality includes data acquisition, quality assessment and data visualization. Starr provides new functions for high level data analysis, e.g., association of ChIP signals with annotated features, gene filtering, and the combined analysis of the ChIP signals and other data like gene expression measurements. It uses the standard data structures for microarray analyses in Bioconductor, building on and fully exploiting the package \Ringo \cite{ringopackage}. The latter implements algorithms for smoothing and peak-finding, as well as low level analysis functions for microarray platforms such as Nimblegen and Agilent.

\section*{Results and discussion}
We demonstrate the utility of \Starr by applying it to a yeast RNA-Polymerase II (PolII for short) ChIP experiment. We discuss the question whether constitutive mRNA expression is mainly determined by the PolII recruitment rate to the promoter.

\subsection*{Data acquisition, quality assessment and normalization}
We facilitated data import as much as possible, since in our experience, this is a major obstacle for the widespread use of R packages in the field of ChIP-chip analysis. The import of data from the microarray manufacturers Nimblegen and Agilent has already been implemented in \Ringo, the common array platform Affymetrix is covered by \Starr. There are two kinds of files that must be known to \Starr: the .bpmap file which contains the mapping of the reporter sequences to its physical position on the array and the .cel files which contain the actual measurement values. All data, no matter from which platform, are stored in the common Bioconductor object {\em ExpressionSet}, which makes them accessible to a number of algorithms operating on that data structure. 
An R script reproducing the entire results of this paper, together with the data stored as RData objects can be found in the supplements. ChIP-chip data of yeast PolII binding was published by Venters and Pugh in 2009 \cite{chipchipdata} and is available on array express under the accession number E-MEXP-1676. The gene expression data used here is available under accession number E-MEXP-2123. Transcription start and termination sites were obtained from David et al. \cite{transcripts}.

The obligatory second step in the analysis protocol is quality control. The complex experimental procedures of a ChIP-chip assay make errors almost inevitable. A special issue of Affymetrix oligo arrays is the bias caused by the GC-content of the oligomer probes \cite{sequenceDependent}. \Starr displays the average expression of probes as a function of their GC-content, and it calculates a position-specific bias of every nucleotide in each of the 25 positions within the probe (see Figure 1).
Moreover, \Starr provides many other quality control plots like an in silico reconstruction of the physical array image to identify flawed regions on the array, or pairwise MA-plots, boxplots and heat-scatter plots to visualize pairwise dependencies within the dataset. 

For the purpose of bias removal (normalization), \Starr interfaces the packages {\em limma} and {\em rMAT}, the latter of which implements the MAT algorithm \cite{MAT}. But it also contains proper normalization methods like the median-rank-percentile normalization, which was originally proposed by Buck and Lieb in 2004 \cite{rankpercentile}.

\subsection*{Visualization and high-level analysis} 
\Starr provides functions for the visualization of a set of ``profiles'' (e.g. time series, signal levels along genomic positions). Figure 2 shows the
ChIP profile of PolII along the transcription start site of genes whose mRNA expression according to \cite{expressiondata} ranges in the least 20\% resp. the top 10\% of all yeast genes (the cutoffs were chosen such that within both groups, the number of genes having an annotated transcription start site was roughly the same). The common way of looking at the intensity profiles is to calculate and plot the mean intensity at each available position along the region of concern. Such an illustration however may hide more than it reveals, since it fails to capture the variability at each position. It is desirable to display this variability in order to assess whether a seemingly obvious alteration in DNA binding is significant or not. Accordingly, our {\em profileplot} function relates to the conventional mean value plot like a box plot relates to an individual sample mean: Let the profiles be given as the rows of a samples $\times$ positions matrix that contains the respective signal of a sample at a given position.
Instead of plotting a line for each profile (e.g. column of the row), the q-quantiles for each position (e.g. column of the matrix) are calculated, where q runs through a set of representative quantiles. Then for each q, the profile line of the q-quantiles is plotted. Color coding of the quantile profiles aids the interpretation of the plot: There is a color gradient from the median profile to the 0 (=min) resp. 1 (=max) quantile.
Another useful high-level plot in \Starr is the {\em correlationPlot}, which displays the correlation of a gene-related binding signal to its corresponding gene expression. Figure 3 shows a plot in which the mean PolII occupancy in various transcript regions of 2526 genes is compared to the corresponding mRNA expression. Each region is defined by its begin and end position relative to the transcription start site (start sites are taken from \cite{transcripts}). The regions are plotted in the lower panel of Figure 3. For each region, the correlation between the vector of mean occupancies and the vector of gene expression values is calculated and shown in the upper panel. 

\subsection*{Results interpretation} 
Figs 2 and 3 supply ambiguous evidence for the role of PolII recruitment in basal transcription: The profile plots suggest that a high PolII occupancy at the initiation region of a gene is a necessary prerequisite for a high mRNA expression level. As opposed to this, the correlation plot reveals that PolII occcupancy at the transcription start is not a good predictor of mRNA expression, but the mean occupancy of PolII in the elongation phase (region 4 in Fig.3) is. Nevertheless, a more detailed analysis of particular gene groups, and a comparison of PolII profiles under different environmental conditions might yield valuable new insights. 

\section*{Conclusion}
\Starr is a Bioconductor package for the analysis of ChIP-chip experiments, in particular of Affymetrix tiling arrays. It exploits the full functionality of \Ringo for the analysis of Affymetrix tiling arrays. These include functions like peak finding, smoothing or plotting genomic regions. \Starr adds new analysis and visualization methods, which can also be applied to two-color technologies. It utilizes standard Bioconductor object classes and can thus easily interface other Bioconductor packages. All functions and methods in the package are well documented in help pages and in a vignette, which illustrates a workflow by means of some example data. Support is provided by the bioconductor mailing list and the package maintainer. \\
Altogehter, \Starr in conjunction with \Ringo constitute a powerful and comprehensive tool for the analysis of tiling arrays across established one- and two-color technologies like Affymetrix, Agilent and Nimblegen.

\section*{Availability and requirements}
The R-package \Starr is available from the Bioconductor
web site at http://www.bioconductor.org and runs on
Linux, Mac OS and MS-Windows. It requires an installed
version of R (version $>=$ 2.10.0), which is freely available
from the Comprehensive R Archive Network (CRAN) at
http://cran.r-project.org, and other Bioconductor packages,
namely Ringo, affy, affxparser, rMAT and vsn plus
the CRAN package pspline and MASS. The easiest way to obtain
the most recent version of the software, with all its
dependencies, is to follow the instructions at http://
www.bioconductor.org/download. \Starr is distributed
under the terms of the Artistic License 2.0.
    
\section*{Authors' contributions}
BZ implemented the \Starr package and did the analysis. AT initiated and supervised the project. Both authors wrote the manuscript and approved of its final version.

\section*{Acknowledgements}
  \ifthenelse{\boolean{publ}}{\small}{}
 We thank Michael Lidschreiber, Andreas Mayer, Matthias Siebert, Johannes Soeding and Kemal Akman for useful comments on the package, Joern Toedling for help on \Ringo, and Anna Ratcliffe for proofreading.


{\ifthenelse{\boolean{publ}}{\footnotesize}{\small}
 \bibliographystyle{bmc_article}  
  \bibliography{bmc_article} }     


\begin{thebibliography}{10}
\providecommand{\url}[1]{[#1]}
\providecommand{\urlprefix}{}

\bibitem{chipchip}
Ren B, Robert F, Wyrick JJ, Aparicio O, Jennings EG, Simon I, Zeitlinger J,
  Schreiber J, Hannett N, Kanin E, Volkert TL, Wilson CJ, Bell SP, Young RA:
  \textbf{Genome-wide location and function of DNA binding proteins.}
  \emph{Science} 2000, \textbf{290}(5500):2306--2309.

\bibitem{bioconductor}
Gentleman RC, Carey VJ, Bates DM, Bolstad B, Dettling M, Dudoit S, Ellis B,
  Gautier L, Ge Y, Gentry J, Hornik K, Hothorn T, Huber W, Iacus S, Irizarry R,
  Leisch F, Li C, Maechler M, Rossini AJ, Sawitzki G, Smith C, Smyth G, Tierney
  L, Yang JY, Zhang J: \textbf{Bioconductor: open software development for
  computational biology and bioinformatics.} \emph{Genome Biol} 2004,
  \textbf{5}(10):R80.

\bibitem{R}
Ihaka R, Gentleman R: \textbf{R: a language for data analysis and graphics.}
  \emph{Journal of Computational and Graphical Statistics} 1996,
  \textbf{5}:299--314.

\bibitem{ringopackage}
Toedling J, Skylar O, Krueger T, Fischer JJ, Sperling S, Huber W:
  \textbf{Ringo--an R/Bioconductor package for analyzing ChIP-chip readouts.}
  \emph{BMC Bioinformatics} 2007, \textbf{8}:221.

\bibitem{chipchipdata}
Venters BJ, Pugh BF: \textbf{A canonical promoter organization of the
  transcription machinery and its regulators in the Saccharomyces genome.}
  \emph{Genome Res} 2009, \textbf{19}(3):360--371.

\bibitem{transcripts}
David L, Huber W, Granovskaia M, Toedling J, Palm CJ, Bofkin L, Jones T, Davis
  RW, Steinmetz LM: \textbf{A high-resolution map of transcription in the yeast
  genome.} \emph{Proc Natl Acad Sci U S A} 2006, \textbf{103}(14):5320--5325.

\bibitem{sequenceDependent}
Royce TE, Rozowsky JS, Gerstein MB: \textbf{Assessing the need for
  sequence-based normalization in tiling microarray experiments.}
  \emph{Bioinformatics} 2007, \textbf{23}(8):988--997.

\bibitem{MAT}
Johnson WE, Li W, Meyer CA, Gottardo R, Carroll JS, Brown M, Liu XS:
  \textbf{Model-based analysis of tiling-arrays for ChIP-chip.} \emph{Proc Natl
  Acad Sci U S A} 2006, \textbf{103}(33):12457--12462.

\bibitem{rankpercentile}
Buck MJ, Lieb JD: \textbf{ChIP-chip: considerations for the design, analysis,
  and application of genome-wide chromatin immunoprecipitation experiments.}
  \emph{Genomics} 2004, \textbf{83}(3):349--360.

\bibitem{expressiondata}
Dengl S, Mayer A, Sun M, Cramer P: \textbf{Structure and in vivo requirement of
  the yeast Spt6 SH2 domain.} \emph{J Mol Biol} 2009, \textbf{389}:211--225.

\end{thebibliography}

\newcommand{\BMCxmlcomment}[1]{}

\BMCxmlcomment{

<refgrp>

<bibl id="B1">
  <title><p>Genome-wide location and function of DNA binding
  proteins.</p></title>
  <aug>
    <au><snm>Ren</snm><fnm>B.</fnm></au>
    <au><snm>Robert</snm><fnm>F.</fnm></au>
    <au><snm>Wyrick</snm><fnm>J. J.</fnm></au>
    <au><snm>Aparicio</snm><fnm>O.</fnm></au>
    <au><snm>Jennings</snm><fnm>E. G.</fnm></au>
    <au><snm>Simon</snm><fnm>I.</fnm></au>
    <au><snm>Zeitlinger</snm><fnm>J.</fnm></au>
    <au><snm>Schreiber</snm><fnm>J.</fnm></au>
    <au><snm>Hannett</snm><fnm>N.</fnm></au>
    <au><snm>Kanin</snm><fnm>E.</fnm></au>
    <au><snm>Volkert</snm><fnm>T. L.</fnm></au>
    <au><snm>Wilson</snm><fnm>C. J.</fnm></au>
    <au><snm>Bell</snm><fnm>S. P.</fnm></au>
    <au><snm>Young</snm><fnm>R. A.</fnm></au>
  </aug>
  <source>Science</source>
  <pubdate>2000</pubdate>
  <volume>290</volume>
  <issue>5500</issue>
  <fpage>2306</fpage>
  <lpage>-2309</lpage>
</bibl>

<bibl id="B2">
  <title><p>Bioconductor: open software development for computational biology
  and bioinformatics.</p></title>
  <aug>
    <au><snm>Gentleman</snm><fnm>R. C.</fnm></au>
    <au><snm>Carey</snm><fnm>V. J.</fnm></au>
    <au><snm>Bates</snm><fnm>D. M.</fnm></au>
    <au><snm>Bolstad</snm><fnm>B.</fnm></au>
    <au><snm>Dettling</snm><fnm>M.</fnm></au>
    <au><snm>Dudoit</snm><fnm>S.</fnm></au>
    <au><snm>Ellis</snm><fnm>B.</fnm></au>
    <au><snm>Gautier</snm><fnm>L.</fnm></au>
    <au><snm>Ge</snm><fnm>Y.</fnm></au>
    <au><snm>Gentry</snm><fnm>J.</fnm></au>
    <au><snm>Hornik</snm><fnm>K.</fnm></au>
    <au><snm>Hothorn</snm><fnm>T.</fnm></au>
    <au><snm>Huber</snm><fnm>W.</fnm></au>
    <au><snm>Iacus</snm><fnm>S.</fnm></au>
    <au><snm>Irizarry</snm><fnm>R.</fnm></au>
    <au><snm>Leisch</snm><fnm>F.</fnm></au>
    <au><snm>Li</snm><fnm>C.</fnm></au>
    <au><snm>Maechler</snm><fnm>M.</fnm></au>
    <au><snm>Rossini</snm><fnm>A. J.</fnm></au>
    <au><snm>Sawitzki</snm><fnm>G.</fnm></au>
    <au><snm>Smith</snm><fnm>C.</fnm></au>
    <au><snm>Smyth</snm><fnm>G.</fnm></au>
    <au><snm>Tierney</snm><fnm>L.</fnm></au>
    <au><snm>Yang</snm><fnm>J. Y.</fnm></au>
    <au><snm>Zhang</snm><fnm>J.</fnm></au>
  </aug>
  <source>Genome Biol</source>
  <pubdate>2004</pubdate>
  <volume>5</volume>
  <issue>10</issue>
  <fpage>R80</fpage>
</bibl>

<bibl id="B3">
  <title><p>R: a language for data analysis and graphics.</p></title>
  <aug>
    <au><snm>Ihaka</snm><fnm>R.</fnm></au>
    <au><snm>Gentleman</snm><fnm>R.</fnm></au>
  </aug>
  <source>Journal of Computational and Graphical Statistics</source>
  <pubdate>1996</pubdate>
  <volume>5</volume>
  <fpage>299</fpage>
  <lpage>-314</lpage>
</bibl>

<bibl id="B4">
  <title><p>Ringo--an R/Bioconductor package for analyzing ChIP-chip
  readouts.</p></title>
  <aug>
    <au><snm>Toedling</snm><fnm>J.</fnm></au>
    <au><snm>Skylar</snm><fnm>O.</fnm></au>
    <au><snm>Krueger</snm><fnm>T.</fnm></au>
    <au><snm>Fischer</snm><fnm>J. J.</fnm></au>
    <au><snm>Sperling</snm><fnm>S.</fnm></au>
    <au><snm>Huber</snm><fnm>W.</fnm></au>
  </aug>
  <source>BMC Bioinformatics</source>
  <pubdate>2007</pubdate>
  <volume>8</volume>
  <fpage>221</fpage>
</bibl>

<bibl id="B5">
  <title><p>A canonical promoter organization of the transcription machinery
  and its regulators in the Saccharomyces genome.</p></title>
  <aug>
    <au><snm>Venters</snm><fnm>B. J.</fnm></au>
    <au><snm>Pugh</snm><fnm>B. F.</fnm></au>
  </aug>
  <source>Genome Res</source>
  <pubdate>2009</pubdate>
  <volume>19</volume>
  <issue>3</issue>
  <fpage>360</fpage>
  <lpage>-371</lpage>
</bibl>

<bibl id="B6">
  <title><p>A high-resolution map of transcription in the yeast
  genome.</p></title>
  <aug>
    <au><snm>David</snm><fnm>L.</fnm></au>
    <au><snm>Huber</snm><fnm>W.</fnm></au>
    <au><snm>Granovskaia</snm><fnm>M.</fnm></au>
    <au><snm>Toedling</snm><fnm>J.</fnm></au>
    <au><snm>Palm</snm><fnm>C. J.</fnm></au>
    <au><snm>Bofkin</snm><fnm>L.</fnm></au>
    <au><snm>Jones</snm><fnm>T.</fnm></au>
    <au><snm>Davis</snm><fnm>R. W.</fnm></au>
    <au><snm>Steinmetz</snm><fnm>L. M.</fnm></au>
  </aug>
  <source>Proc Natl Acad Sci U S A</source>
  <pubdate>2006</pubdate>
  <volume>103</volume>
  <issue>14</issue>
  <fpage>5320</fpage>
  <lpage>-5325</lpage>
</bibl>

<bibl id="B7">
  <title><p>Assessing the need for sequence-based normalization in tiling
  microarray experiments.</p></title>
  <aug>
    <au><snm>Royce</snm><fnm>T. E.</fnm></au>
    <au><snm>Rozowsky</snm><fnm>J. S.</fnm></au>
    <au><snm>Gerstein</snm><fnm>M. B.</fnm></au>
  </aug>
  <source>Bioinformatics</source>
  <pubdate>2007</pubdate>
  <volume>23</volume>
  <issue>8</issue>
  <fpage>988</fpage>
  <lpage>-997</lpage>
</bibl>

<bibl id="B8">
  <title><p>Model-based analysis of tiling-arrays for ChIP-chip.</p></title>
  <aug>
    <au><snm>Johnson</snm><fnm>W. E.</fnm></au>
    <au><snm>Li</snm><fnm>W.</fnm></au>
    <au><snm>Meyer</snm><fnm>C. A.</fnm></au>
    <au><snm>Gottardo</snm><fnm>R.</fnm></au>
    <au><snm>Carroll</snm><fnm>J. S.</fnm></au>
    <au><snm>Brown</snm><fnm>M.</fnm></au>
    <au><snm>Liu</snm><fnm>X. S.</fnm></au>
  </aug>
  <source>Proc Natl Acad Sci U S A</source>
  <pubdate>2006</pubdate>
  <volume>103</volume>
  <issue>33</issue>
  <fpage>12457</fpage>
  <lpage>-12462</lpage>
</bibl>

<bibl id="B9">
  <title><p>ChIP-chip: considerations for the design, analysis, and application
  of genome-wide chromatin immunoprecipitation experiments.</p></title>
  <aug>
    <au><snm>Buck</snm><fnm>M. J.</fnm></au>
    <au><snm>Lieb</snm><fnm>J. D.</fnm></au>
  </aug>
  <source>Genomics</source>
  <pubdate>2004</pubdate>
  <volume>83</volume>
  <issue>3</issue>
  <fpage>349</fpage>
  <lpage>-360</lpage>
</bibl>

<bibl id="B10">
  <title><p>Structure and in vivo requirement of the yeast Spt6 SH2
  domain.</p></title>
  <aug>
    <au><snm>Dengl</snm><fnm>S.</fnm></au>
    <au><snm>Mayer</snm><fnm>A.</fnm></au>
    <au><snm>Sun</snm><fnm>M.</fnm></au>
    <au><snm>Cramer</snm><fnm>P.</fnm></au>
  </aug>
  <source>J Mol Biol</source>
  <pubdate>2009</pubdate>
  <volume>389</volume>
  <issue>1</issue>
  <fpage>211</fpage>
  <lpage>-225</lpage>
</bibl>

</refgrp>
} 


\ifthenelse{\boolean{publ}}{\end{multicols}}{}



\section*{Figures}
  \subsection*{Figure 1 - Hybridization bias}
     Sequence-specific dependency of raw reporter intensities. (A) Boxplots of probe intensity distributions. Probes are grouped according to the C/C content in their sequence. The median intensity 
     increases with rising G/C-content. (B) Position-dependent mean probe intensity. Each letter corresponds the mean intensity of all probes that contain the corresponding nucleotide in the respective position.

 \subsection*{Figure 2 - PolII along the transcriptional start site}
  	Profiles of PolII occupancy of genes with low (least $20\%$) resp high (top $10\%$) transcription rates (cluster 1 resp. cluster 2).
	The upper graphs show the mean occupancy calculated over each position along the transcription start site. The lower plots illustrate the variance in the two clusters. The black line indicates the median profile of all features. The color gradient corresponds to quantiles (from 0.05 to 0.95), and the first and third quartiles are shown as grey lines. The light grey lines in the background show the profiles of	individual "outlier" features.

 \subsection*{Figure 3 - Correlation of PolII occupancy to gene expression}
      \Starr enables the systematic investigation of gene expression related to DNA binding. Figure 2 shows the correlation of
     the mean PolII occupancy within different regions along the transcript to gene expression. 
     The lower panel shows the regions of interest relative to the transcription start site (TSS) and the transcription termination site (TTS).
     The upper panel shows the correlation of PolII occupancy to the gene expression of the corresponding regions.

\end{bmcformat}
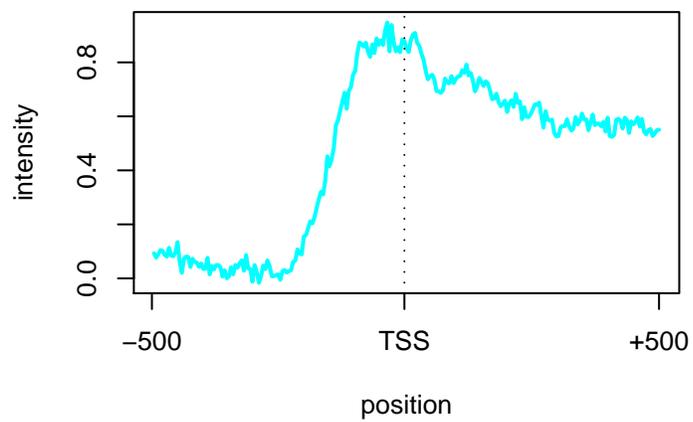
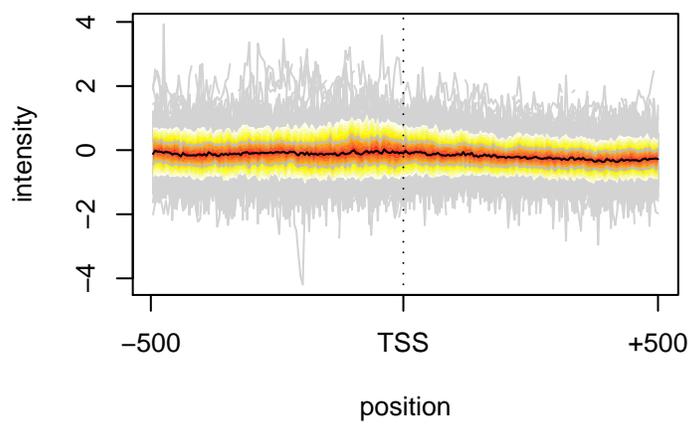
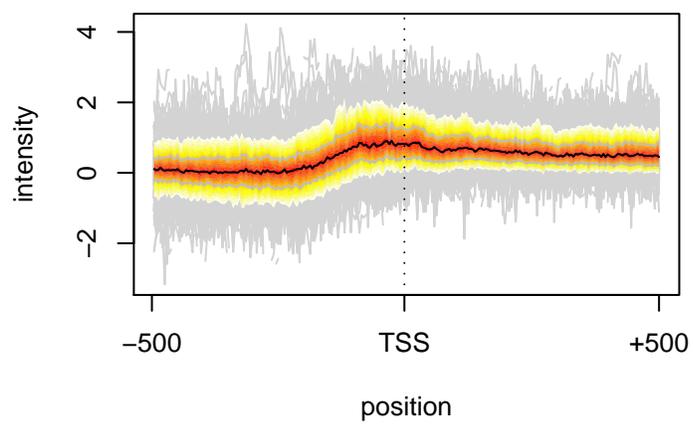
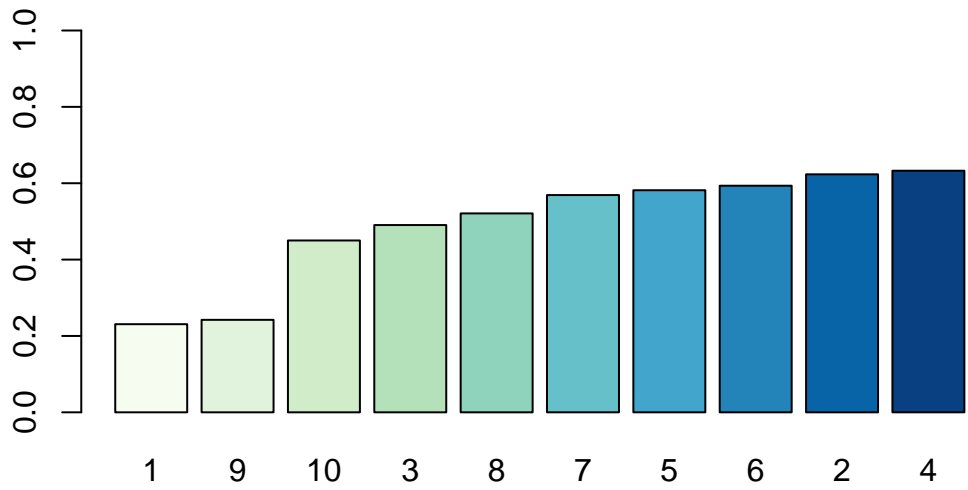
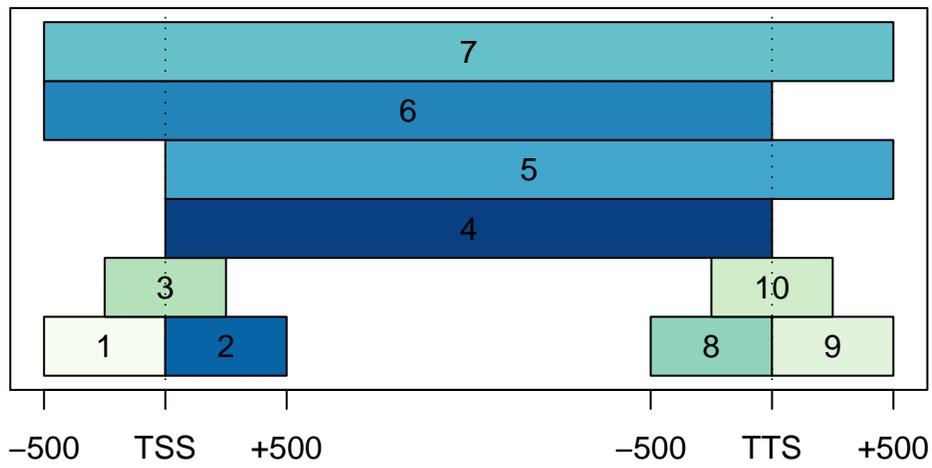
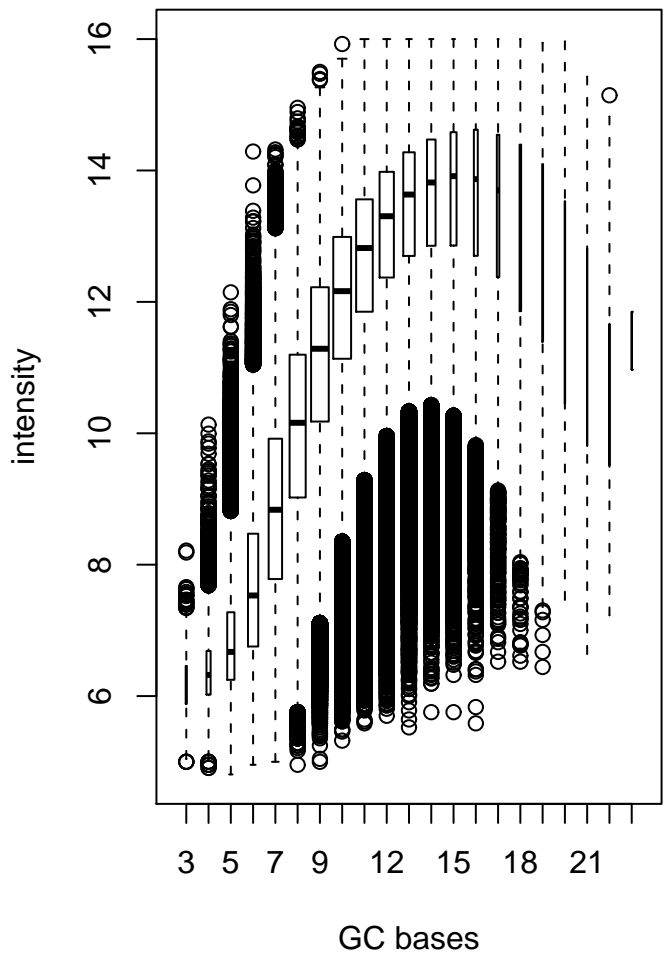
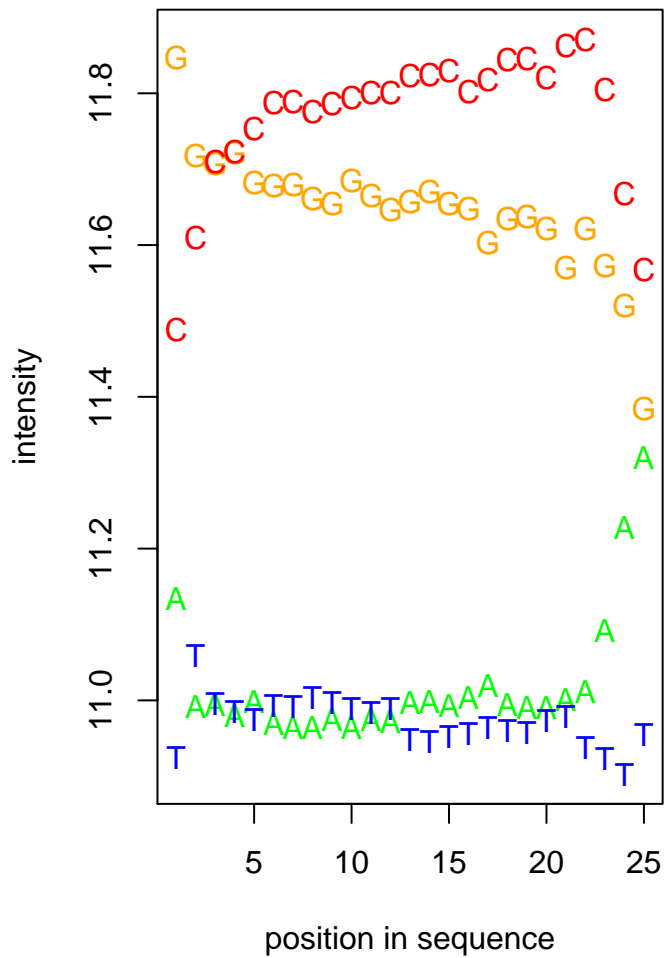
\end{document}